\documentclass[letterpaper, 12pt, twoside, notitlepage]{article}
\pdfoutput=1
\usepackage[margin=1in]{geometry}
\usepackage[singlespacing]{setspace}
\usepackage{cite}
\usepackage{amsthm, amsmath, amssymb}
\usepackage{url}
\usepackage{syntonly}
\usepackage{textcomp}
\usepackage{graphicx}
\usepackage{comment}
\usepackage[font={small}, labelfont=bf]{caption}
\usepackage{hyperref}
\usepackage{subcaption}
\usepackage{amstext}
\usepackage{array}
\usepackage{caption}

\title{\fontsize{19}{20}\selectfont \textbf{Glacial Cycles and Milankovitch Forcing}}
\date{}
\author{\fontsize{15}{16}\selectfont Shiv Priyam Raghuraman\\ \fontsize{13}{14}\selectfont School of Mathematics, University of Minnesota, Minneapolis, MN, USA}

\begin{document}
\maketitle

\begin{abstract}
Using a recent conceptual model of the glacial-interglacial cycles we present more evidence of Milankovitch cycles being the trigger for retreat and forming of ice sheets in the cycles. This model is based on a finite approximation of an infinite dimensional model which has three components: Budyko's energy balance model describing the annual mean temperatures at latitudes, Widiasih's ODE which describes the behavior of the edge of the ice sheet, and Walsh et al. who introduced a snow line to account for glacial accumulation and ablation zones. Certain variables in the model are made to depend on the Milankovitch cycles, in particular, the obliquity of the Earth's axis and the eccentricity of the Earth's orbit. We see as a result that deglaciation and glaciation do occur mostly due to obliquity and to some extent eccentricity.
\end{abstract}

\section{\fontsize{12}{13}\selectfont Introduction}
Over the last one million years, massive ice sheets across North America have periodically formed through Earth's history during glacial periods. Following this cold period which lasts about 90,000 years, ice sheets melt relatively quickly, in what is called the interglacial period which is characterized by warmer temperatures and lasts about 10,000 years. Subsequently after this 100,000 year cycle, the ice sheets slowly form again in the glacial period and the cycle progresses. 

Milankovitch hypothesized that variations in the Earth's orbital parameters affect the amount of incoming solar radiation the Earth's surface receives. Variations in the following orbital parameters are postulated to pace the glacial cycles: obliquity of the Earth's spin axis, i.e., the tilt of the Earth's axis, eccentricity of the Earth's orbit, and the precession of the Earth as it rotates, i.e., the amount it ``wobbles". However, orbital forcing cannot be the only reason behind the pacing of the glacial cycles, \cite{hays1976variations} and so, there must be nonlinear feedbacks inherent to Earth's climate system. \cite{walsh}

Conceptual climate models are used to investigate a small number of these feedbacks at a time since they are computationally less intensive as compared to general circulation models. One such feedback is the ice-albedo feedback, which is modeled as a dynamical system \cite{mcgehee2012paleoclimate}. Ice-albedo feedback is a positive feedback climate process where a change in the area of snow-covered land, ice caps, glaciers or sea ice alters the albedo, i.e., the ratio of reflected radiation to the incident radiation. Cooling tends to increase ice cover and hence the albedo, reducing the amount of solar energy absorbed and leading to more cooling. Conversely, warming tends to decrease ice cover and hence the albedo, which increases the amount of solar energy absorbed, leading to more warming \cite{deser2000arctic}. 

Budyko was interested in how ice-albedo feedback affects climate and in his 1969 paper \cite{budyko1969effect}, he introduces the conceptual energy balance model (EBM). We build our mathematical model of glacial cycles on Budyko's model and use a quadratic approximation by McGehee and Widiasih \cite{mcgehee2014quadratic} to reduce the infinite dimensional system to a pair of ordinary differential equations by introducing a variable called the ice line which indicates the edge of the ice sheet. By adding another variable called the snow line \cite{walsh}, which is independent of the ice line, we aimed to generate glacial cycles using this simple model. We were able to do so and observe that the variations in obliquity of Earth's orbit affects the glacial cycles predominantly and eccentricity of Earth's orbit plays a role as well.

\section{\fontsize{12}{13}\selectfont Temperature-Ice line model}
We look at Budyko's model which studies the average annual temperatures in latitudinal zones. Consider Budyko's time-dependent equation \cite{held1974simple}:\\
\begin{equation}
R\frac{\partial T(y,t)}{\partial t} = Qs(y)(1-\alpha (y))-(A+BT) - C\left(T-\overline{T}\right),
\end{equation}
This equation represents the change in energy stored in the Earth's surface at $y\in [0,1]$, where $y$ is the sine of the latitude with $y=0$ the equator and $y=1$ the north pole since Budyko's EBM assumes the world to be symmetric about the equator. $T=T(y,t) (^{\circ}\mathrm{C})$ is the annual average surface temperature on the circle of latitude at $y$. The units of each side of (1) are Watts per meter squared $\left(\frac{\text{W}}{\text{m}^{2}}\right)$. The quantity $R$ is the specific heat of the Earth's surface, measured in units of $\frac{\text{J}}{\text{m}^{2}\,^{\circ}\mathrm{C}}$.
$Q$ represents the average annual incoming solar radiation (also known as insolation), a parameter which depends on the eccentricity of Earth's orbit \cite{mcgehee2012paleoclimate}. $s(y)$ depends upon the obliquity of Earth's orbit \cite{mcgehee2012paleoclimate}, which describes the distribution of insolation across a latitude, and satisfies $\int_{0}^{1} s(y) \mathrm{d}y=1$. $\alpha_{y}$ denotes the planetary albedo, which as described earlier, measures the extent to which insolation is reflected back into space. Thus, the first term on the right hand side of (1) represents the energy absorbed at latitude $y$ on the surface from the sun. 

The energy reradiated into space at longer wavelengths is approximated linearly by the term $A+BT$. Before the heat escapes into space, some of it is absorbed by greenhouse gases and returned to the surface. Thus, this reradiation term is the net loss of energy from the surface to space. The energy transported from warmer latitudes to cooler latitudes is approximated by the term $C(\overline{T}-T)$ where $\overline{T}$ is the global annual average surface temperature and satisfies $\overline{T}=\int_{0}^{1}T(y,t) \mathrm{d}y$. $A,B, \text{and} \, C$ are positive constants found empirically through satellite data. \cite{mcgehee2014quadratic}

The equilibrium temperature profiles are found to be: \cite{mcgehee2014quadratic}
\begin{equation}
T^{*}(y)=\frac{1}{B+C}\left(Qs(y)(1-\alpha(y))-A+\frac{C}{B}(Q(1-\overline{\alpha}-A)\right),
\end{equation}
where
\begin{equation}
\overline{\alpha}=\int_{0}^{1} \alpha(y)s(y)\mathrm{d}y.
\end{equation}
\begin{equation}
s(y)=s_{0}p_{0}(y)+s_{2}p_{2}(y), s_{0}=1, s_{2}=-0.482, 
\end{equation}
where $p_{0}(y)=1$ and $p_{2}(y)=\frac{1}{2}(3y^2-1)$ are the first two even Legendre polynomials. (4) is within 2\% of the true $s(y)$ values \cite{north1975theory}.

A key feature of Budyko's model is that it assumes that the Earth has an ice cap, with the requirement that above a particular latitude $y=\eta$ there is always ice and below the latitude $y=\eta$ there's no ice. The ice line is then defined to be the edge of the ice sheet $\eta$. Consequently, we can write the albedo function as:
\begin{equation}
\alpha_{\eta}(y) = \left\{
\begin{array}{lr}
\alpha_{1}, & \text{if}\, y<\eta\\
\alpha_{2}, & \text{if}\, y>\eta,\\
\end{array}
\right.
\end{equation}
where $\alpha_{1}<\alpha_{2}$ and $\alpha_{1}$ denotes the albedo of the surface having no ice and $\alpha_{2}$ denotes the albedo of the surface having ice. Using (4) and (5), the equilibrium temperature profiles (2) are even, piecewise quadratic functions having a discontinuity at $\eta$. Note that $\eta$ parametrizes (3) and hence (2), so we write $T_{\eta}^{*}(y)$. Therefore, for each value of $\eta$ there are infinitely many equilibrium temperature functions. We define $T^{*}(\eta)$ as $T^{*}(\eta)=\frac{\lim_{y\to\eta^{-}} T^{*}(y)+\lim_{y\to\eta^{+}} T^{*}(y)}{2}$ and the equilibrium temperature at the ice line as
\begin{equation}
T_{\eta}^{*}(\eta)=\frac{1}{B+C}\left(Qs(\eta)(1-\alpha_{0})-A+\frac{C}{B}(Q(1-\overline{\alpha}-A)\right),
\end{equation}
where $\alpha_{0}=\frac{\alpha_{1}+\alpha_{2}}{2}$. In particular, Budyko was interested in finding out for which values of $\eta$ does $T_{\eta}^{*}(\eta)=T_c$, where $T_c$ is a critical temperature above which ice melts and below which ice forms. 
However, Budyko's model does not permit the ice line $\eta$ to respond to changes in temperature. This drawback was solved by Widiasih in \cite{widiasih2013dynamics} where an ODE modeling the evolution of $\eta$ was added, giving the following system:
\begin{subequations}
\begin{align}
R\frac{\partial T}{\partial t} &= Qs(y)(1-\alpha (y,\eta))-(A+BT) - C\left(T-\overline{T}\right)\\
\frac{d\eta}{dt}&=\rho (T(\eta,t)-T_c),
\end{align}
\end{subequations}
where $\rho$ is a parameter which controls the relaxation time of the ice sheet. The above system describes how the temperature distribution $T(y,t)$ evolves according to Budyko's equation (7a) and the evolution of $\eta$ in (7b). If $T(\eta,t)>T_c$, the ice sheets melt and retreat toward the pole. If $T(\eta,t)<T_c$, the ice sheets expand and move toward the equator.\cite{walsh}

\section{\fontsize{12}{13}\selectfont Approximation to finite dimensional system}
As mentioned earlier, the equilibrium solutions of Budyko's equation (7a) are even and piecewise quadratic, with a discontinuity at $\eta$ when using (4) and (5). This prompted the introduction of a quadratic approximation to the infinite dimensional system (7), which we will summarize below while the reader can find all details in \cite{mcgehee2014quadratic}. Let $X$ denote the space of even, piecewise quadratic functions having a discontinuity at $\eta$. The four-dimensional linear space $X$ can be parameterized by the new variables $w_0,z_0,w_2, \text{and}\, z_2$ by letting
\begin{equation}
T(y) = \left\{
\begin{array}{lr}
w_0+\frac{z_0}{2}+\left(w_2+\frac{z_2}{2}p_{2}(y)\right), &y<\eta\\
w_0-\frac{z_0}{2}+\left(w_2-\frac{z_2}{2}p_{2}(y)\right), &y>\eta\\
w_0+w_{2}p_{2}(\eta), &y=\eta.\\
\end{array}
\right.
\end{equation}
$T(\eta)$ above is consistent with (6) and $T(\eta)=\frac{\lim_{y\to\eta^{-}} T^{*}(y)+\lim_{y\to\eta^{+}} T^{*}(y)}{2}$.
\begin{equation}
\overline{T}=\int_{0}^{1}T(y)\mathrm{d}y=w_0+z_{0}\left(\eta-\frac{1}{2}+z_{2}P_{2}(\eta)\right), \text{where} \, P_{2}(\eta)=\int_{0}^{\eta}p_{2}(y)\mathrm{d}y.
\end{equation}
In fact, $P_{2}(\eta)=\int_{0}^{\eta}p_{2}(y)\mathrm{d}y=\int_{0}^{\eta}\frac{3y^2-1}{2}\mathrm{d}y=\frac{\eta^3-\eta}{2}$.
Plugging (8) and (4) into (7a) and equating the coefficients of $p_{0}(y)$, i.e., the constant terms and $p_{2}(y)$ respectively, yields the following equations, two each for $y<\eta$ and $y>\eta$:
\begin{subequations}
\begin{align}
R\left(\dot{w_{0}}+\frac{\dot{z_{0}}}{2}\right) &= Q(1-\alpha_1)-A-(B+C)\left(w_0+\frac{z_0}{2}\right)+C\overline{T}\\
R\left(\dot{w_{0}}-\frac{\dot{z_{0}}}{2}\right) &= Q(1-\alpha_2)-A-(B+C)\left(w_0-\frac{z_0}{2}\right)+C\overline{T}\\
R\left(\dot{w_{2}}+\frac{\dot{z_{2}}}{2}\right) &= Qs_{2}(1-\alpha_1)-(B+C)\left(w_2+\frac{z_2}{2}\right)\\
R\left(\dot{w_{2}}-\frac{\dot{z_{2}}}{2}\right) &= Qs_{2}(1-\alpha_2)-(B+C)\left(w_2-\frac{z_2}{2}\right)
\end{align}
\end{subequations}
Adding and subtracting equations (10a)-(10b) and (10c)-(10d) and plugging in (9) for $\overline{T}$, one gets:
\begin{subequations}
\begin{align}
R\dot{w_{0}} &= Q(1-\alpha_0)-A-Bw_0+C\left(\left(\eta-\frac{1}{2}\right)z_0+z_{2}P_{2}(\eta)\right)\\
R\dot{z_{0}} &= Q(\alpha_2-\alpha_1)-(B+C)z_0\\
R\dot{w_{2}} &= Qs_{2}(1-\alpha_0)-(B+C)w_2\\
R\dot{z_{2}} &= Qs_{2}(\alpha_2-\alpha_1)-(B+C)z_2.
\end{align}
\end{subequations}
Let $L=\frac{Q}{B+C}$. Notice that system (11) admits a globally attracting invariant line $l$ and one can prove that on $l$, system (11) reduces to the single equation which provides the approximation to (7a)
\begin{equation}
R\dot{w}=-B(w-F(\eta)),
\end{equation}
where for the sake of convenience $w=w_0$ and where $F(\eta)$ is a cubic polynomial
\begin{equation}
F(\eta)=\frac{1}{B}\left(Q(1-\alpha_0)\-A+CL(\alpha_2-\alpha_1)\left(\eta-\frac{1}{2}+s_{2}P_{2}(\eta)\right)\right).
\end{equation}
One computes the expression for $T(\eta)$ on the invariant line $l$ which provides the approximation for (7b) as seen in \cite{mcgehee2014quadratic}
\begin{equation*}
\dot{\eta}=\rho(w-G(\eta))
\end{equation*}
where 
\begin{equation}
G(\eta)=-Ls_{2}(1-\alpha_{0})p_{2}(\eta)+T_c,
\end{equation}
when restricting to $l$. Thus, the infinite dimensional system (7) is approximated by the system of ODEs
\begin{subequations}
\begin{align}
\dot{w}&=-\tau(w-F(\eta))\\
\dot{\eta}&=\rho(w-G(\eta)),
\end{align}
\end{subequations}
where $F(\eta), G(\eta)$ are given in (13) and (14) respectively, and $\tau=\frac{B}{R}$. One can prove that for fixed $\eta$ the variable $w$ is a translate of the global average temperature (intuitively similar to $\overline{T}$, but different because $w$ is biased by the latitude). In \cite{mcgehee2014quadratic}, it is proven that there exists a stable equilibrium point with a small ice cap, and a saddle equilibrium point with a large ice cap, for all $\rho>0$, for standard parameter values (See Section 5). \cite{walsh}

However, since $\eta$ approaches either a small ice cap or the equator over time, it doesn't take into account the relative sizes of the accumulation and ablation (melting) zones in glacial advance and retreat and hence does not permit glacial cycles. In \cite{walsh}, this was rectified by adding a variable called the snow line independent of the ice line, which is presented below as well.

\section{\fontsize{12}{13}\selectfont Snow line addition}
The accumulation and ablation of ice play a fundamental role in the theory of glacial cycles, serving to control the terminus advance and retreat, the ice volume, and the geometry of the surface of the ice sheet \cite{bahr1997physical}. Abe-Ouchi et al in \cite{abe2013insolation} found the fast retreat of the ice sheet was due to significantly enhanced ablation, i.e., the ablation rate for a large, advancing ice sheet was necessarily much smaller than the ablation rate for a retreating ice sheet, in order to faithfully reproduce the last four glacial cycles. \cite{walsh} This simple idea was incorporated into the model below. Another motivation for constructing the model below comes from the ``flip-flop" model of the thermohaline circulation in \cite{welander1982simple}. The model vector field has a line of discontinuity that produces a switch to the alternate regime. The model below will share some similarities with \cite{welander1982simple} in this sense.

Considering system (15) again, independent snow and ice lines are introduced to incorporate accumulation and ablation zones. We begin by recasting the role played by $\eta$, interpreting $\eta$ henceforth as the snow line. $\xi$ will denote the (more slowly moving) ice line, i.e., the edge of the ice sheet (see Figure 1 below). The ablation zone has extent $\eta-\xi$ (when $\eta>\xi$), while the accumulation zone has size $1-\eta$. \cite{walsh}
\begin{figure}[h!]
\centering
\includegraphics{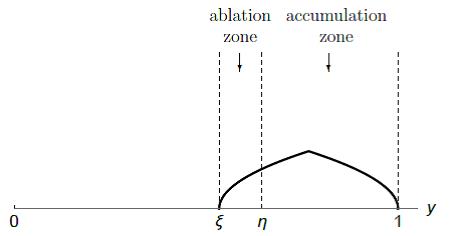}
\caption{Model set-up. $\eta$ is the snow line and $\xi$ is the ice line. For the sake of illustration, the shape presented represents a glacier. Taken from Walsh et al \cite{walsh}, reproduced with permission from the authors.}
\end{figure}
The temperature-ice line-snow line model is a non-smooth system with state space \cite{walsh}
\begin{equation*}
\mathcal{B}=\{(w,\eta,\xi):w\in\mathbb{R}, \eta\in[0,1], \xi\in[0,1]\}
\end{equation*}
defined as follows. Choose parameters $b_0<b<b_1$ denoting ablation rates, and a parameter $a$ representing the accumulation rate. When $b(\eta-\xi)-a(1-\eta)<0$, the ice sheet advances since accumulation exceeds ablation \cite{walsh}, we \textit{set} 
\begin{subequations}
\begin{align}
\dot{w}&=-\tau(w-F(\eta))\\
\dot{\eta}&=\rho(w-G_{-}(\eta))\\
\dot{\xi}&=\epsilon(b_0(\eta-\xi)-a(1-\eta)).
\end{align}
\end{subequations}
$F(\eta)$ in (16a) is given by (13), $G_{-}(\eta)$ in (16b) is given by (14) but with $T_c=T_{c}^{-}=-5 ^{\circ}\mathrm{C}$, and $\epsilon>0$ is a time constant governing the movement of the ice line.\\
When $(b(\eta-\xi)-a(1-\eta)>0$, the ice sheet retreats since ablation exceeds accumulation, we \textit{set}
\begin{subequations}
\begin{align}
\dot{w}&=-\tau(w-F(\eta))\\
\dot{\eta}&=\rho(w-G_{+}(\eta))\\
\dot{\xi}&=\epsilon(b_1(\eta-\xi)-a(1-\eta)).
\end{align}
\end{subequations}
where $F(\eta)$ in (17a) is given by (13), $G_{+}(\eta)$ in (16b) is given by (14) but with $T_c=T_{c}^{+}=-10 ^{\circ}\mathrm{C}$.

The relative sizes of ablation rates $b_0$ and $b_1$ were motivated by Abe-Ouchi’s paper \cite{abe2013insolation}. The choice of different $T_c$-values is motivated by Tziperman’s paper \cite{tziperman2003mid}, in which a linear interpolation between $T_c=-13^{\circ}\mathrm{C}$ and $T_c=-3 ^{\circ}\mathrm{C}$ is introduced to model changes in deep ocean temperature. The idea behind it is that a large ice sheet that is advancing implies a colder world overall, so then less energy is required to form ice (and vice versa for a retreating ice sheet). \cite{walsh}

We thus arrive at a 3-dimensional system having a plane of discontinuity
\begin{equation}
\Sigma=\{(w,\eta,\xi):b(\eta-\xi)-a(1-\eta)=0\}=\left\{(w,\eta,\xi):\xi=\left(1+\frac{a}{b}\right)\eta-\frac{a}{b}\right\}.
\end{equation}
As we will see, a trajectory in $(w,\eta,\xi)$-space passing through $\Sigma$ switches from an advancing state to a state of glacial retreat, or vice versa \cite{walsh}, which is similar in terms of the idea to the flip-flop model in \cite{welander1982simple}.

\section{\fontsize{12}{13}\selectfont Milankovitch forcing}
\subsection{\fontsize{12}{13}\selectfont Equilibrium Solutions}
The model is now forced with Milankovitch cycles, i.e., we make important parameters such as $Q$ and $s_2$ depend on eccentricity of Earth’s orbit and obliquity of Earth's axis respectively. McGehee and Lehman in \cite{mcgehee2012paleoclimate} showed that insolation, $Q$ is a function of $e$, the eccentricity of the Earth's orbit, given by
\begin{equation}
Q=Q(e)=\frac{Q_0}{\sqrt{1-e^2}},
\end{equation}
where $Q_0$ is the insolation assuming the eccentricity of Earth's orbit is $0$ (See Table 1 for value).
McGehee and Widiasih in \cite{mcgehee2012simplification} proved that the function $s_{2}$ in (4) actually depends on $\beta$, the obliquity of Earth's axis, given by
\begin{equation}
s_2=s_2 (\beta)=\frac{5}{16}(−2+3 \sin^2⁡\beta ).
\end{equation}
Since $F(\eta)$ and $G(\eta)$ given in (13) and (14) are functions of $Q$ and $s_2$, this makes $F(\eta)$ and $G(\eta)$ functions of $e \, \text{and} \, \beta$. 
\begin{table}[h!]
\begin{center}
\begin{tabular}{|l|c|r|}
\hline
Parameter & Value & Units\\ \hline
$Q_{0}$ & $343$ & $\text{W m}^{-2}$\\
$A$ & $202$ & $\text{W m}^{-2}$ \\
$B$ & $1.9$ & $\text{W m}^{-2}(^{\circ}\mathrm{C})^{-1}$\\
$C$ & $3.04$ & $\text{W m}^{-2}(^{\circ}\mathrm{C})^{-1}$ \\
$\alpha_1$ & $0.32$ & dimensionless\\
$\alpha_2$ & $0.62$ & dimensionless\\
$T_{c}^{+}$ & $-10$ & $^{\circ}\mathrm{C}$ \\
$T_{c}^{-}$ & $-5.5$ & $^{\circ}\mathrm{C}$ \\ 
\hline
\end{tabular}
\end{center}
\caption{Parameter values (taken from \cite{walsh})}
\end{table}

Note that the accumulation and ablation parameters $a,b,b_0,b_1$ are dimensionless constants, $\tau, \epsilon$ have units $(\text{seconds})^{-1}$, and $\rho$ has units $(\text{seconds})^{-1}(^{\circ}\mathrm{C})^{-1}$.

Consider the system of ODEs in (15) again. 
\begin{equation*}
\begin{split}
\dot{w}&=-\tau(w-F(\eta))\\
\dot{\eta}&=\rho(w-G(\eta)),
\end{split}
\end{equation*}
In order to obtain the equilibrium solutions, we set the derivatives equal to $0$. That is,
\begin{equation*}
w=F(\eta)=G(\eta) \Rightarrow F(\eta)-G(\eta)=0.
\end{equation*}
Recall that $F(\eta)$ is a cubic polynomial in $\eta$ due to $P_{2}(\eta)=\frac{\eta^3-\eta}{2}$ and $G(\eta)$ is a quadratic polynomial in $\eta$ due to $p_{2}(\eta)=\frac{3\eta^2-1}{2}$. Thus, three roots are found when solving the cubic equation $F(\eta)-G(\eta)=0$ and one is discarded since it does not belong to the range $[0,1]$. These roots are the snow line values that are dependent on both $e$ and $\beta$. We aim to find out which factor it is more dependent on. 

We see in Figure 2 that for $T_c =-10^{\circ}\mathrm{C}$, which has a small stable (sink) ice cap near the pole, $\eta$, the snow line, is varying in sync with the obliquity values.
\begin{figure}[h!]
\centering
\includegraphics[scale=0.7]{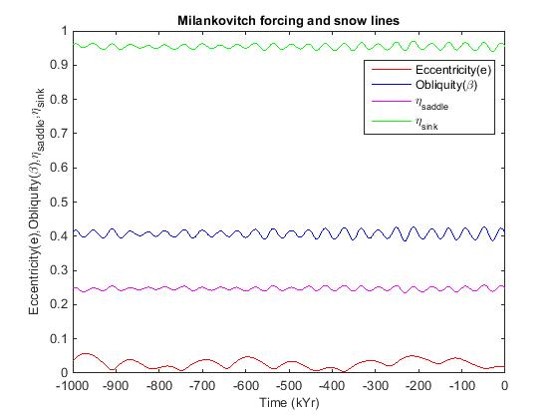}
\caption{Snow-lines affected by Milankovitch cycles when $T_c =-10^{\circ}\mathrm{C}$}
\end{figure}
In Figure 3, we notice that the stable (sink) ice cap, which is larger than the previous case since not as much energy is required to form ice at $T_c =-5.5^{\circ}\mathrm{C}$, doesn't evidently vary only with $\beta$ but instead depends on both $e$ and $\beta$. The unstable (saddle) large ice cap below varies closely with $\beta$. 
\begin{figure}[h!]
\centering
\includegraphics[scale=0.7]{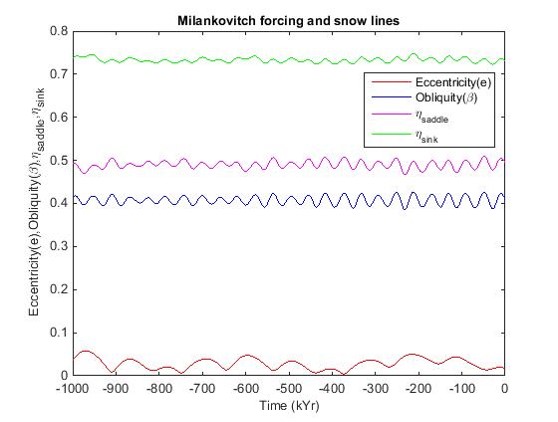}
\caption{Snow-lines affected by Milankovitch cycles when $T_c =-5.5^{\circ}\mathrm{C}$}
\end{figure}
Obliquity of the Earth's axis playing the dominant role in how ice sheets form was also predicted by McGehee and Lehman in \cite{mcgehee2012paleoclimate}. Thus, this model and the above figures lends more weight to their claim. 

\subsection{\fontsize{12}{13}\selectfont Ice line-Snow line dynamics}
We first show that we do get glacial cycles with this model and the similarities to the ``flip-flop" model in \cite{welander1982simple}. Now, that we have $\eta$ we can also calculate $\xi$ using (18) and finally find the discontinuity equation given by the first equality in (18). We begin with $T_c =-10^{\circ}\mathrm{C}$. Since it takes a lot of energy to go lower than $-10^{\circ}\mathrm{C}$ and form large ice sheets, a small ice cap is formed (Figure 4). For $a=1.05, b=1.75, b_1=5, b_0=1.5, D=b(\eta-\xi)-a(1-\eta)$ and sink values of $\eta,\xi$, we get $D<0$ which means that there is more accumulation than ablation. 
\begin{figure}[h!]
\centering
\includegraphics[scale=0.7]{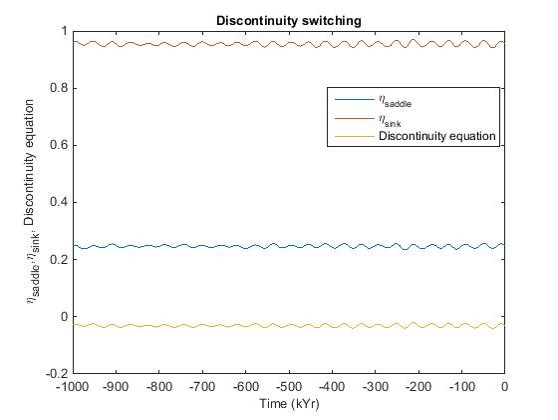}
\caption{Discontinuity equation and switching mechanism when $T_c =-10^{\circ}\mathrm{C}$}
\end{figure}
Recall that when $D<0$, we set $T_c=-5.5^{\circ}\mathrm{C}$ and we switch to the ODEs in system (16) and we have Figure 5.
\begin{figure}[h!]
\centering
\includegraphics[scale=0.7]{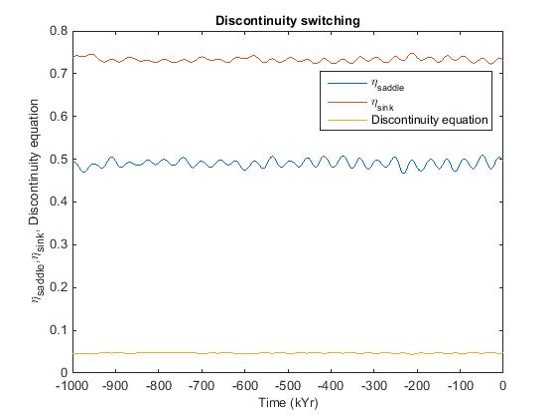}
\caption{Discontinuity equation and switching mechanism when $T_c =-5.5^{\circ}\mathrm{C}$}
\end{figure}
We note that for $T_c=-5.5^{\circ}\mathrm{C}$ we have a larger ice cap since less energy is required to go lower than $-5.5^{\circ}\mathrm{C}$ as compared to $-10^{\circ}\mathrm{C}$. For the same values of $a, b, b_1, b_0$ as above, we get $D>0$, which means that there is more ablation than accumulation. However, recall that when $D>0$, we set $T_c =-10^{\circ}\mathrm{C}$ and switch to the ODEs in system (17), which corresponds exactly to Figure 4 above. But after this, $D<0$ and we swtich back to Figure 5. Thus, we have found for these parameter values, glacial cycles do occur, where glaciation and deglaciation take place one after the other. In particular, although our model is constantly in the ``flip-flop" state, $D$ never crosses 0. If it were 0, accumulation equals ablation and hence there would be no trigger to go to the next state. 

We now make $b=b_0$ to reduce our parameter space.
First, we force the model with two Milankovitch parameters, obliquity and eccentricity. In Figure 6(a) we note that we get cycles, that is, we constantly move from a glaciation state to a deglaciation state and vice versa. When the trajectory crosses the discontinuity place, ablation exceeds accumulation and retreat begins or accumulation exceeds ablation and ice sheets expand toward the equator (larger values of $\xi$). 

Figure 6(b) describes the evolution of the snow line and ice line over time. $t=0$ represents one million years ago and $t=1000$ represents present day. Note that the ice line (red) increases slowly, denoting a slow descent into a long glacial period and an expanding ice sheet and then a short interglacial period follows where the ice sheet relatively quickly retreats as seen in the steep increase of $\xi$. This is consistent with paleoclimate data \cite{petit1999climate}. 

Huybers in \cite{huybers2007glacial} made a case that obliquity must be the trigger for ice sheet retreat since otherwise ice sheets were too massive to melt rapidly without any external factor. Most importantly, as seen in Figure 6(b) deglacial events were not occurring every obliquity cycle and instead skipped two obliquity cycles at multiple times, such as at $t=200,400,800$. Measuring from peak to peak, we see that these skips were between $80$ and $120$ kyr which is exactly what Huybers predicted in \cite{huybers2007glacial}. 

\begin{figure}
\begin{subfigure}[b]{0.5\textwidth}
\includegraphics[width=\textwidth]{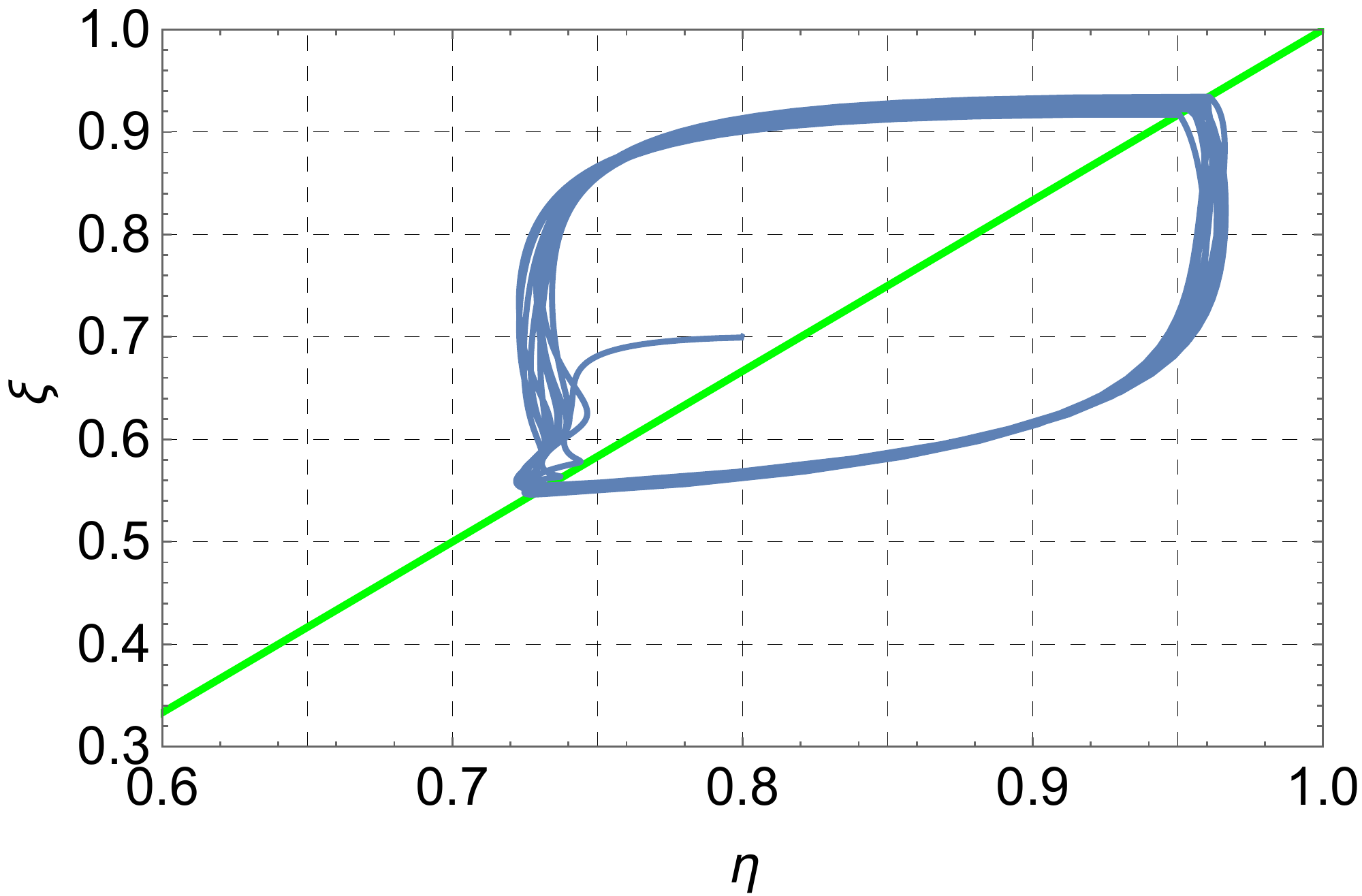}
\caption{Discontinuity plane cycles with full Milankovitch forcing}
\label{fig:1}
\end{subfigure}
\begin{subfigure}[b]{0.5\textwidth}
\includegraphics[width=\textwidth]{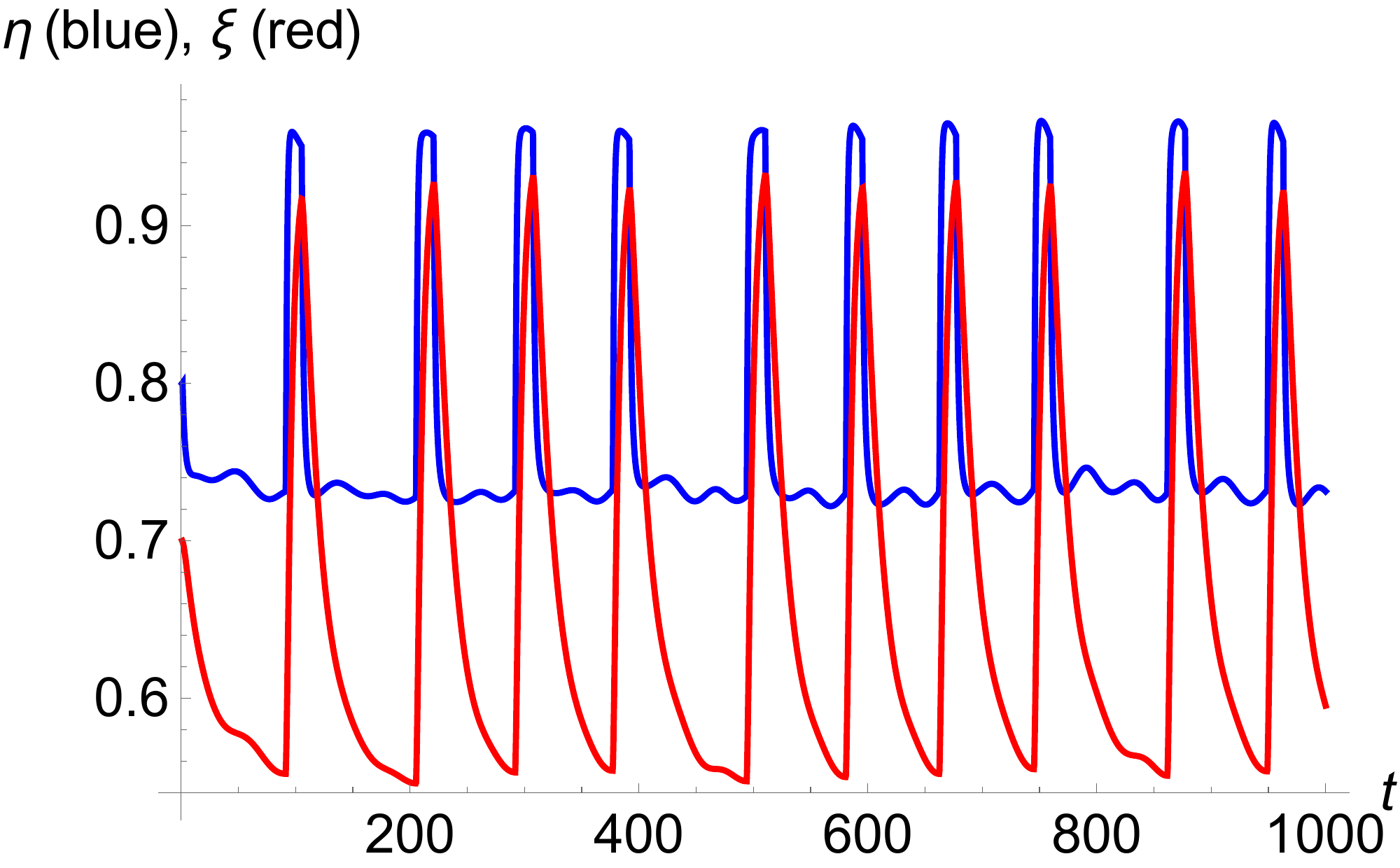}
\caption{Evolution of ice and snow lines over time with full Milankovitch forcing}
\label{fig:2}
\end{subfigure}
\caption{$b=b_0=1.5, b_1=5, a=1, \rho=\epsilon=4\times 10^{-2}$}
\end{figure}

However, when we force our model only with obliquity (Figure 7(b)), that is, we remove variations due to eccentricity by assuming $Q(e)=Q_0$, near $t=800$ there is no obliquity cycle which is skipped. Also note that the skipping of cycles near $t=400$ in Figure 6(b) is pushed back to $t=300$ in Figure 7(b). This leads us to consider the possibility that eccentricity still does play a role in these cycles. In Figure 8(b) we see exactly this, as there are skipping of cycles consistently when obliquity is removed, i.e., $s_2=s_{2}(\beta)$. This is in agreement with Huybers' revised outlook in \cite{huybers2011combined} where he notes that eccentricity is also responsible for deglaciation.

\begin{figure}
\begin{subfigure}[b]{0.5\textwidth}
\includegraphics[width=\textwidth]{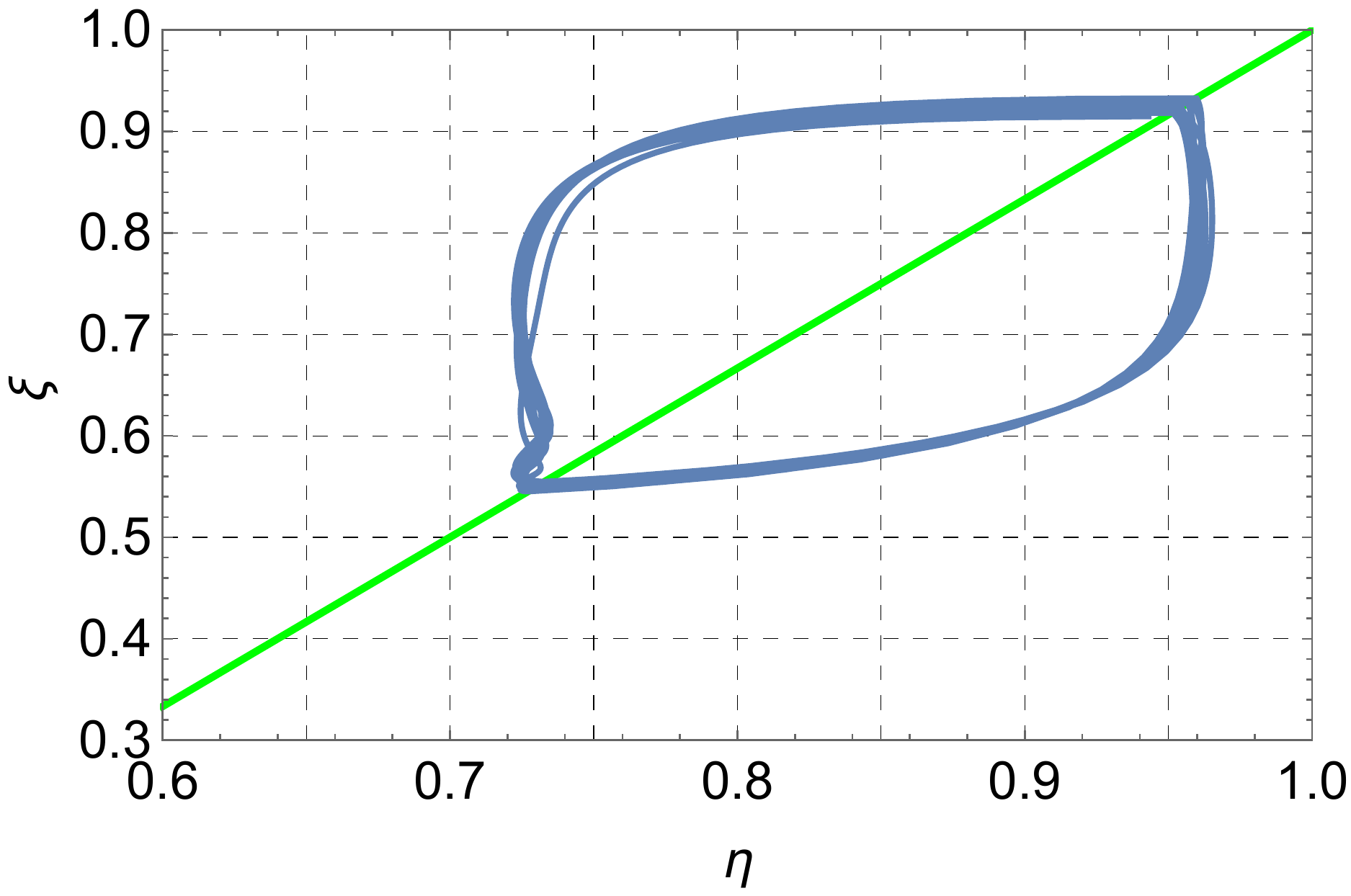}
\caption{Discontinuity plane cycles with oblqiuity only forcing}
\label{fig:3}
\end{subfigure}
\begin{subfigure}[b]{0.5\textwidth}
\includegraphics[width=\textwidth]{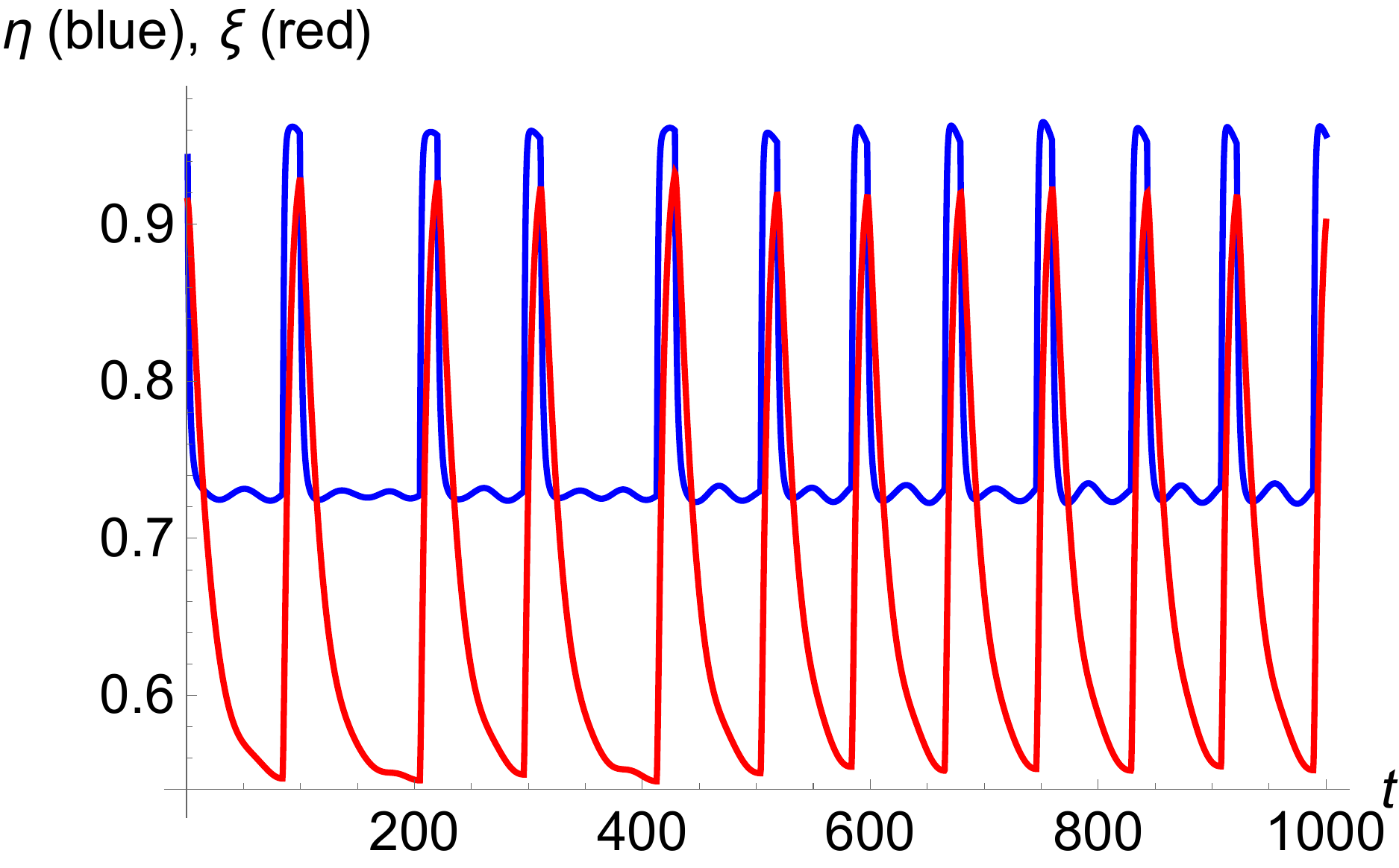}
\caption{Evolution of ice and snow lines over time with obliquity only forcing}
\label{fig:4}
\end{subfigure}
\caption{$b=b_0=1.5, b_1=5, a=1, \rho=\epsilon=4\times 10^{-2}$}
\end{figure}

\begin{figure}
\begin{subfigure}[b]{0.5\textwidth}
\includegraphics[width=\textwidth]{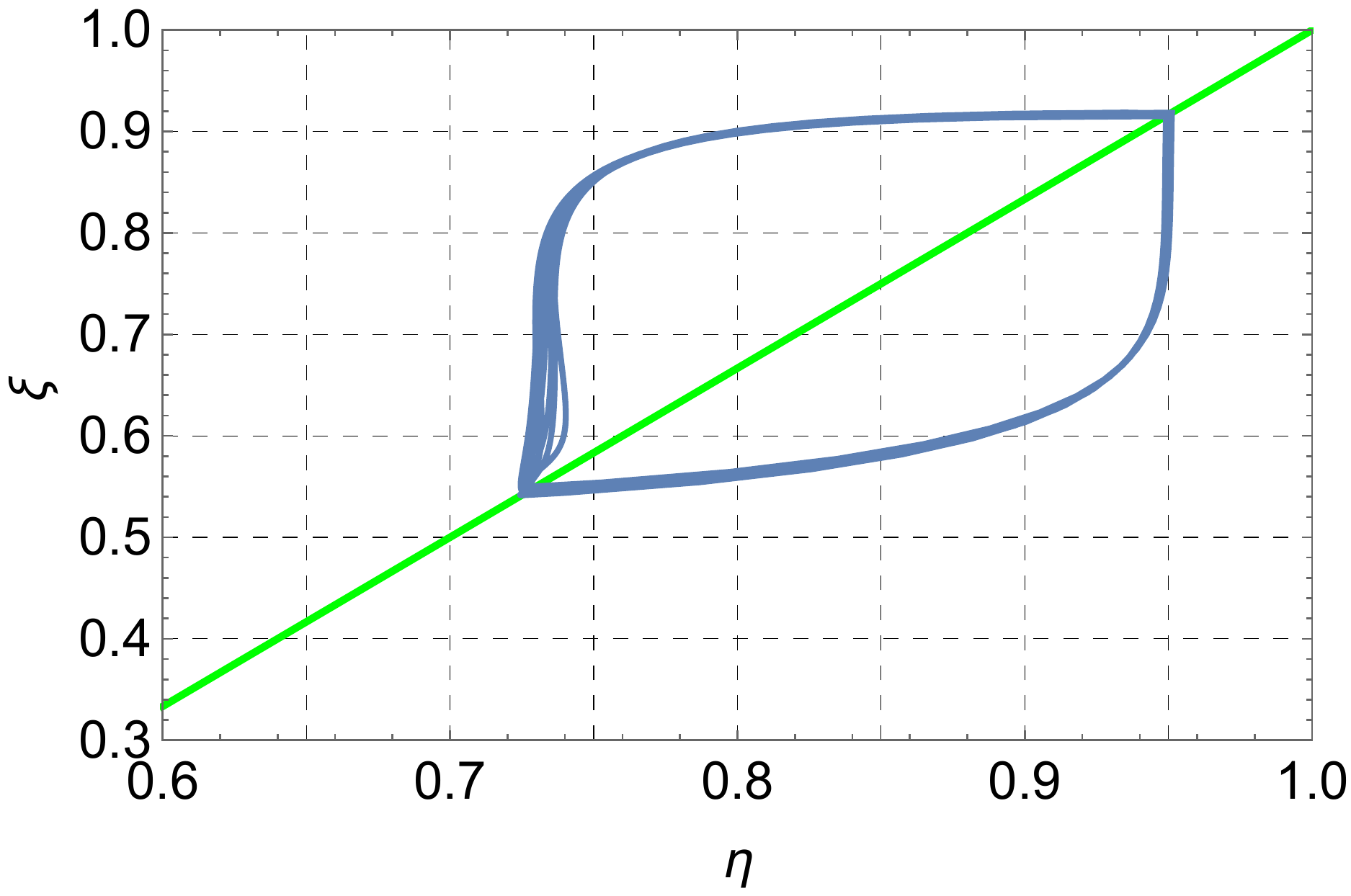}
\caption{Discontinuity plane cycles with eccentricity only forcing}
\label{fig:5}
\end{subfigure}
\begin{subfigure}[b]{0.5\textwidth}
\includegraphics[width=\textwidth]{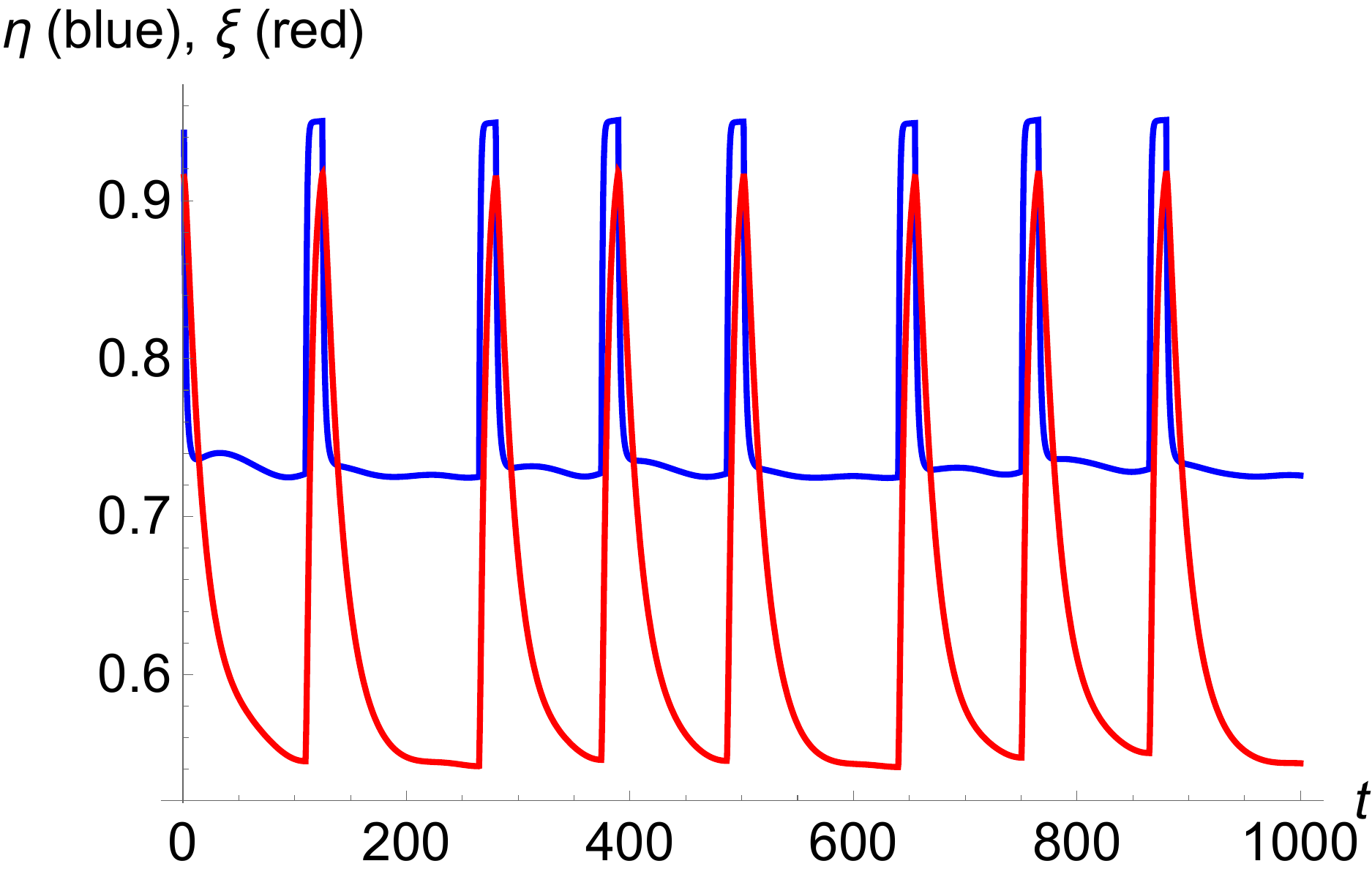}
\caption{Evolution of ice and snow lines over time with eccentricity only forcing}
\label{fig:6}
\end{subfigure}
\caption{$b=b_0=1.5, b_1=5, a=1, \rho=\epsilon=4\times 10^{-2}$}
\end{figure}

In Figures 9,10,11 we make the time constant for the snow line 100 times faster than the time constant for the ice line. Interestingly, we note that there is not much of a change in the evolution of the ice lines and snow lines over time even though the trajectory across the discontinuity plane has changed.
\begin{figure}
\begin{subfigure}[b]{0.5\textwidth}
\includegraphics[width=\textwidth]{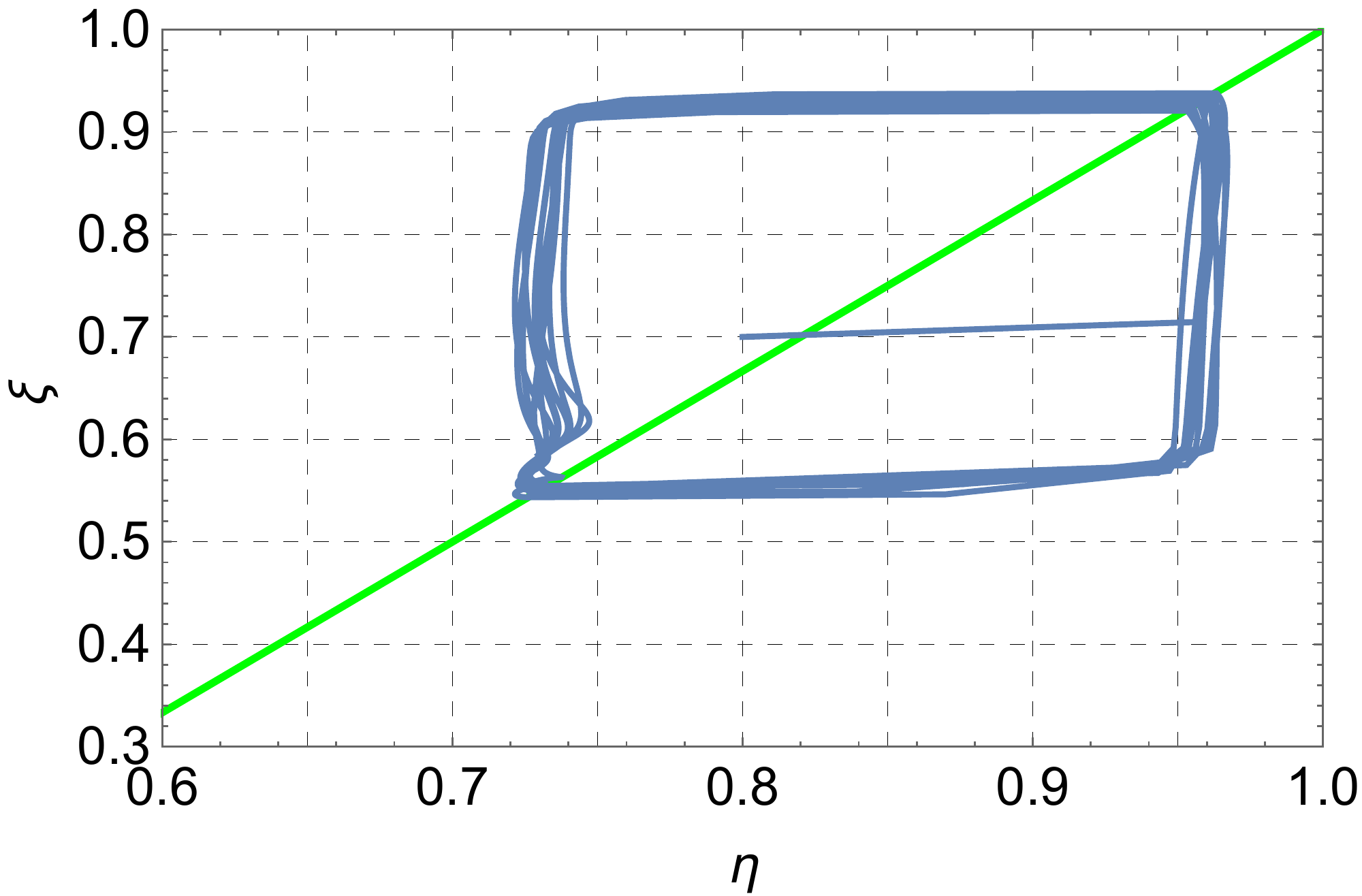}
\caption{Discontinuity plane cycles with full Milankovitch forcing}
\label{fig:7}
\end{subfigure}
\begin{subfigure}[b]{0.5\textwidth}
\includegraphics[width=\textwidth]{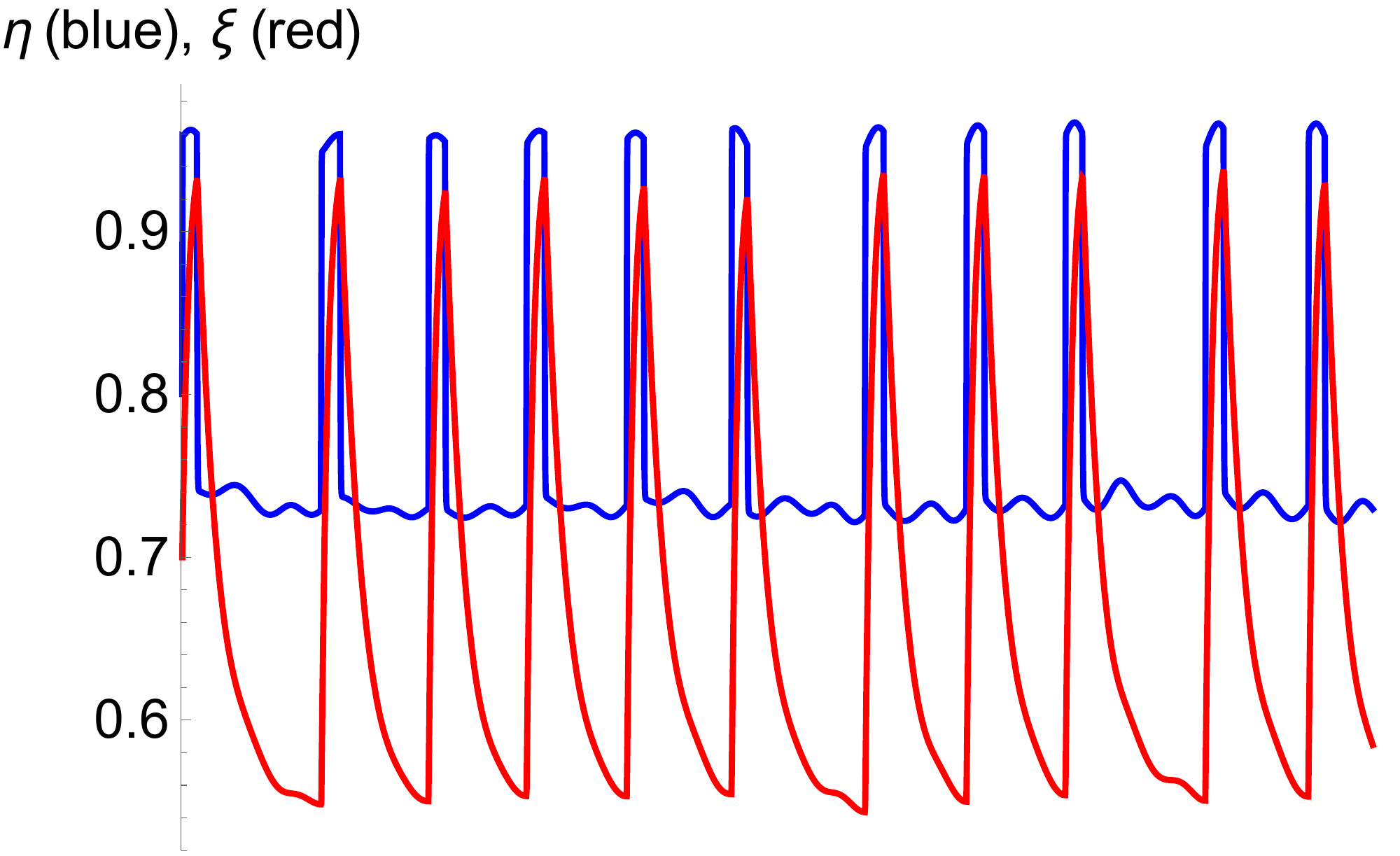}
\caption{Evolution of ice and snow lines over time with full Milankovitch forcing}
\label{fig:8}
\end{subfigure}
\caption{$b=b_0=1.5, b_1=5, a=1, \rho=100\epsilon=4$}
\end{figure}

\begin{figure}
\begin{subfigure}[b]{0.5\textwidth}
\includegraphics[width=\textwidth]{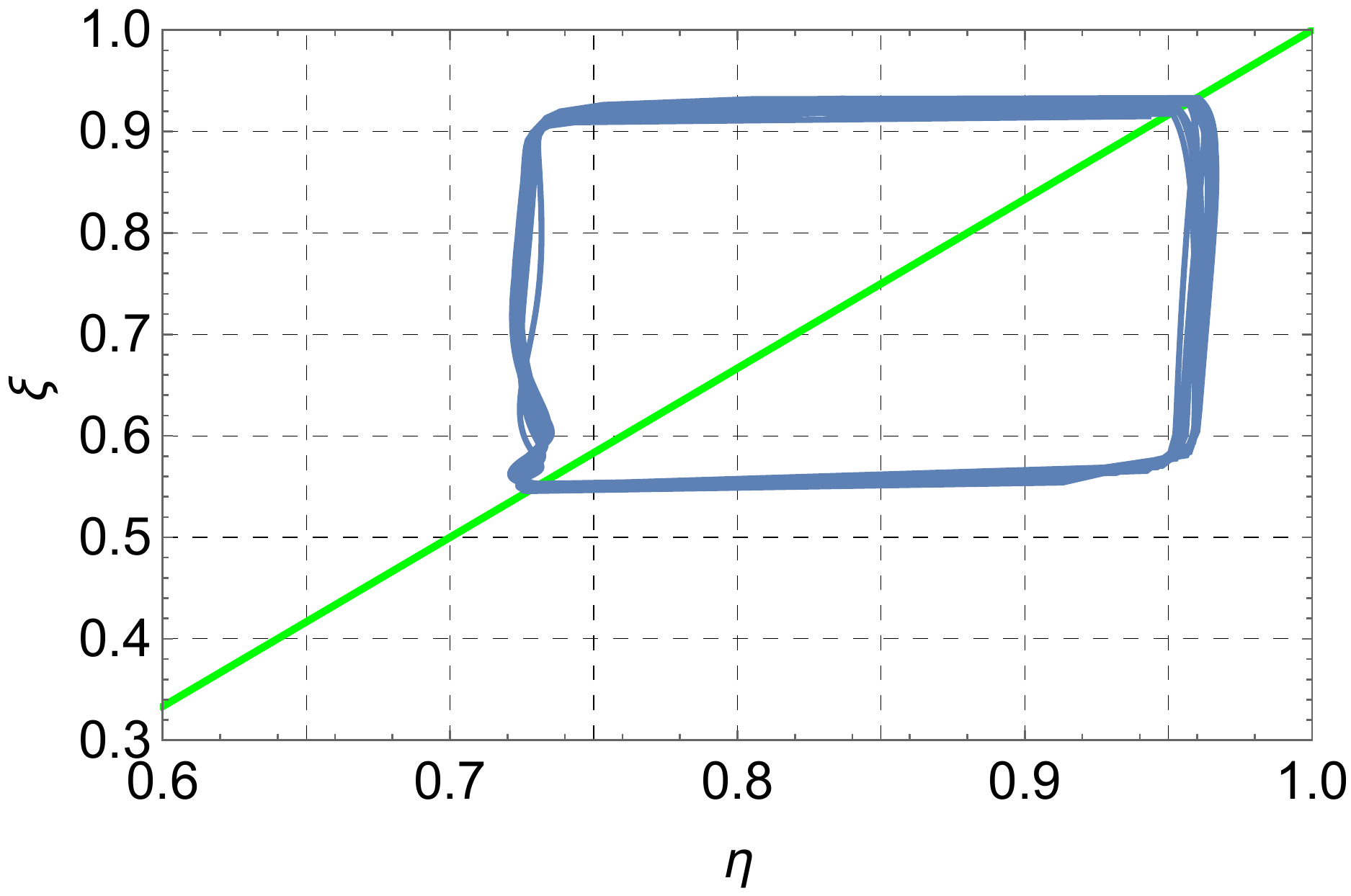}
\caption{Discontinuity plane cycles with obliquity only forcing}
\label{fig:9}
\end{subfigure}
\begin{subfigure}[b]{0.5\textwidth}
\includegraphics[width=\textwidth]{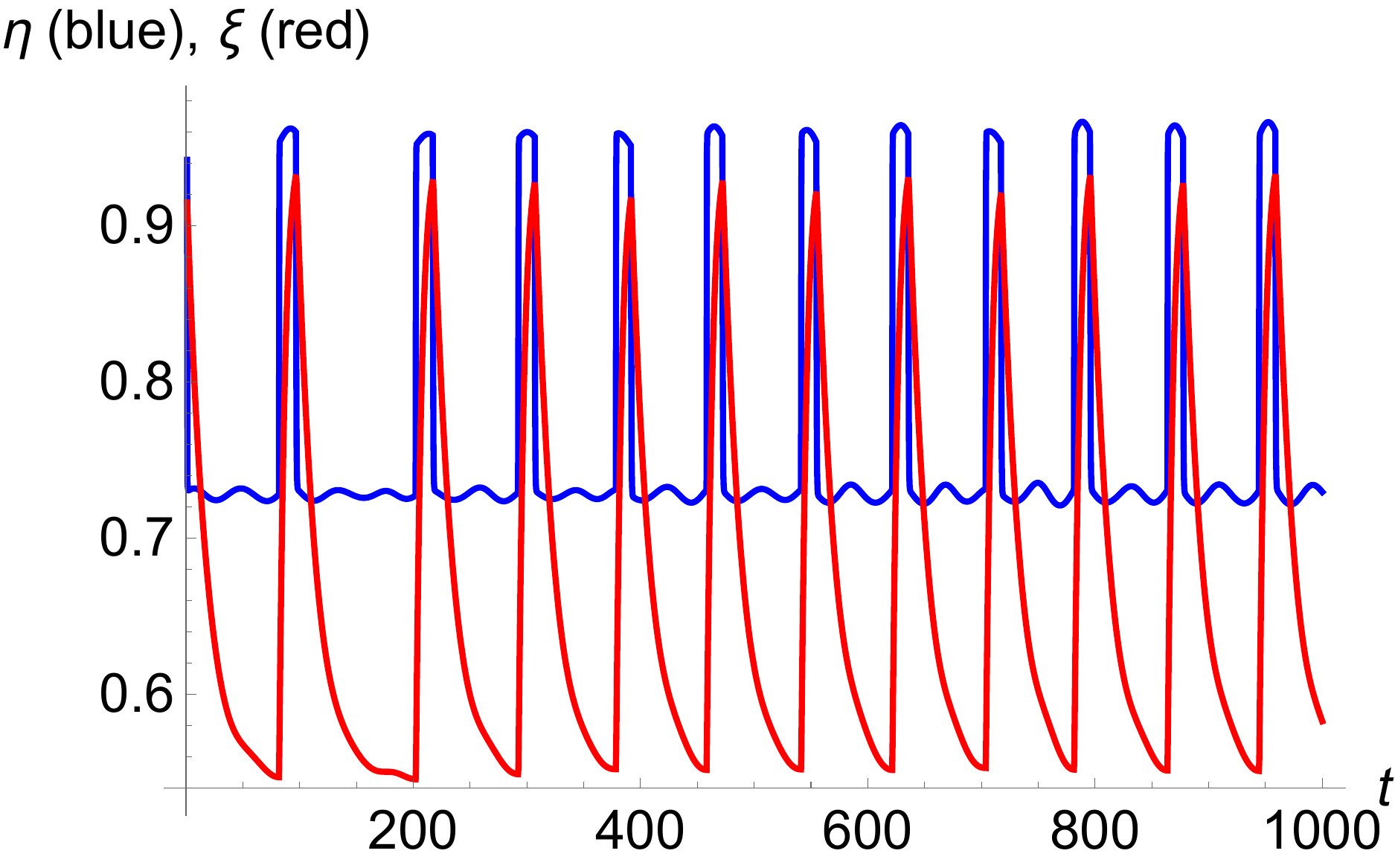}
\caption{Evolution of ice and snow lines over time with full Milankovitch forcing}
\label{fig:10}
\end{subfigure}
\caption{$b=b_0=1.5, b_1=5, a=1, \rho=100\epsilon=4$}
\end{figure}

\begin{figure}
\begin{subfigure}[b]{0.5\textwidth}
\includegraphics[width=\textwidth]{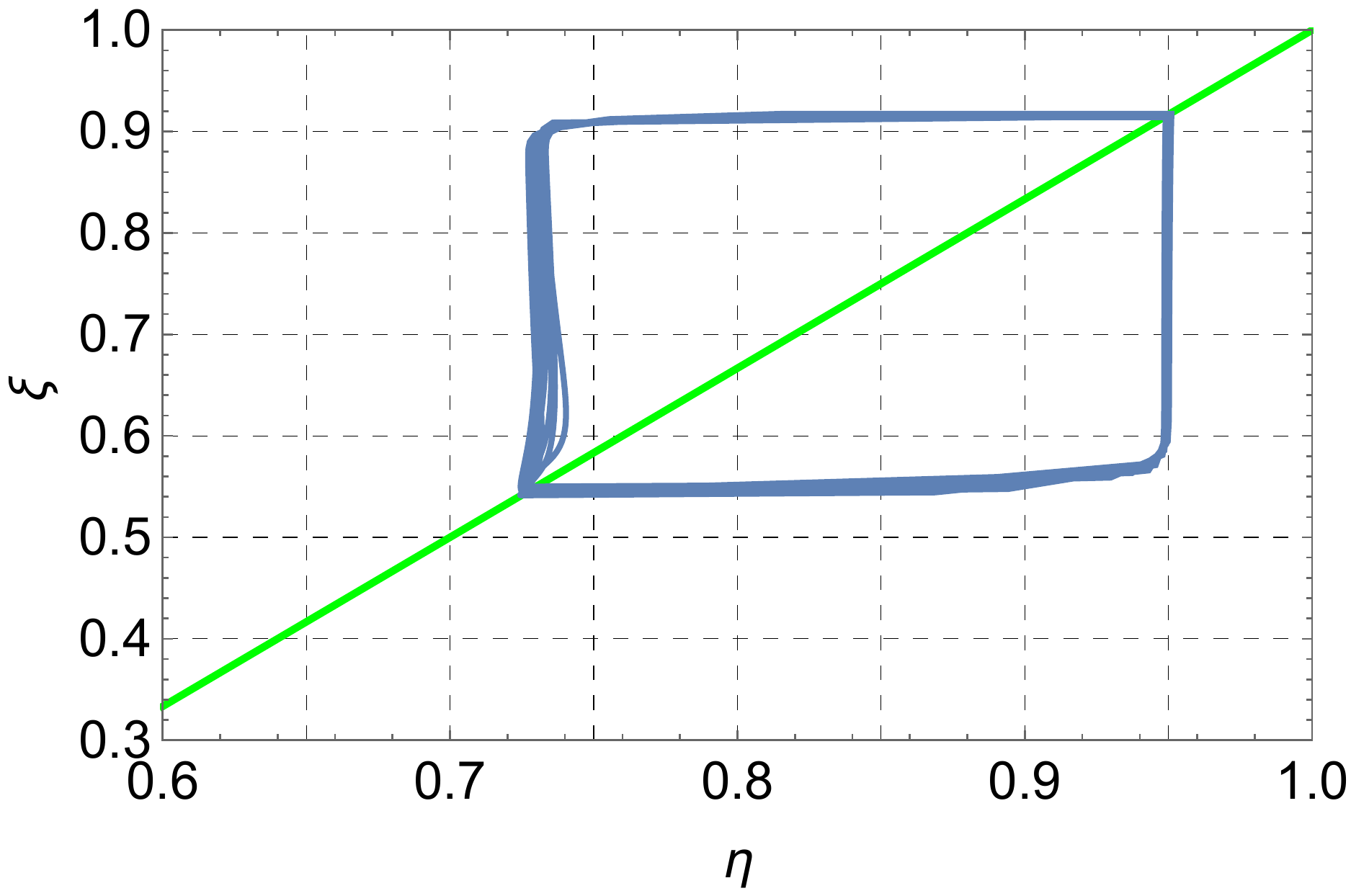}
\caption{Discontinuity plane cycles with eccentricity only forcing}
\label{fig:11}
\end{subfigure}
\begin{subfigure}[b]{0.5\textwidth}
\includegraphics[width=\textwidth]{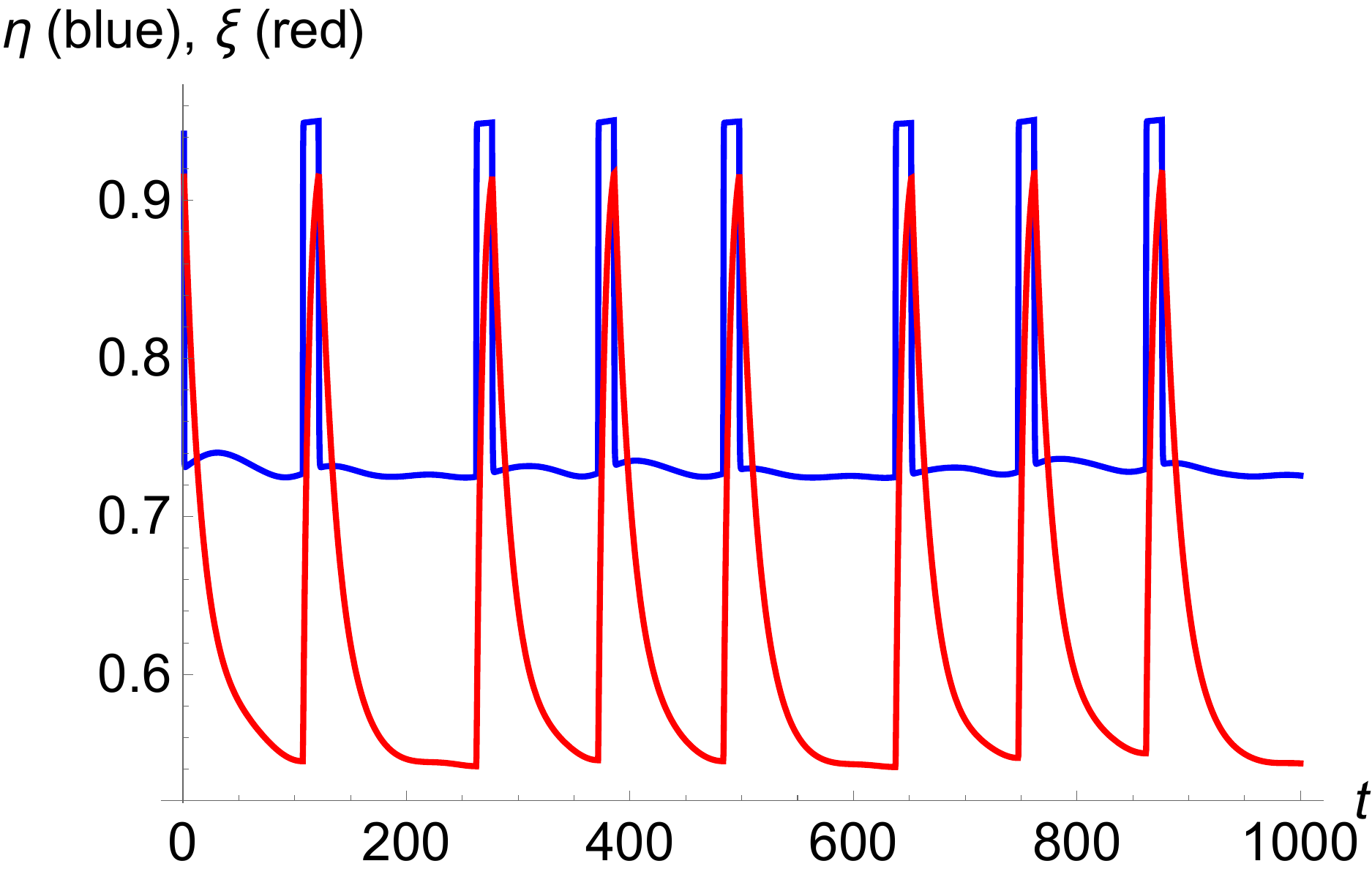}
\caption{Evolution of ice and snow lines over time with eccentricity only forcing}
\label{fig:12}
\end{subfigure}
\caption{$b=b_0=1.5, b_1=5, a=1, \rho=100\epsilon=4$}
\end{figure}

We now increase the ablation parameter $b_1$ to its limiting case $b_1=45$, beyond this value we do not get the expected glacial-interglacial cycles. When we force the system fully with Milankovitch cycles we observe in Figure 12(a) that the trajectory does not spend much time ablating and quickly crosses the discontinuity plane to start accumulating ice slowly. We expect this since $b_1=45$ is a large ablation value and pushes the ice sheets to retreat faster. In Figure 12(b) we see the rapid interglacial period and slow descent into the glacial period as well and still have the skipping of some obliquity cycles. 

\begin{figure}
\begin{subfigure}[b]{0.5\textwidth}
\includegraphics[width=\textwidth]{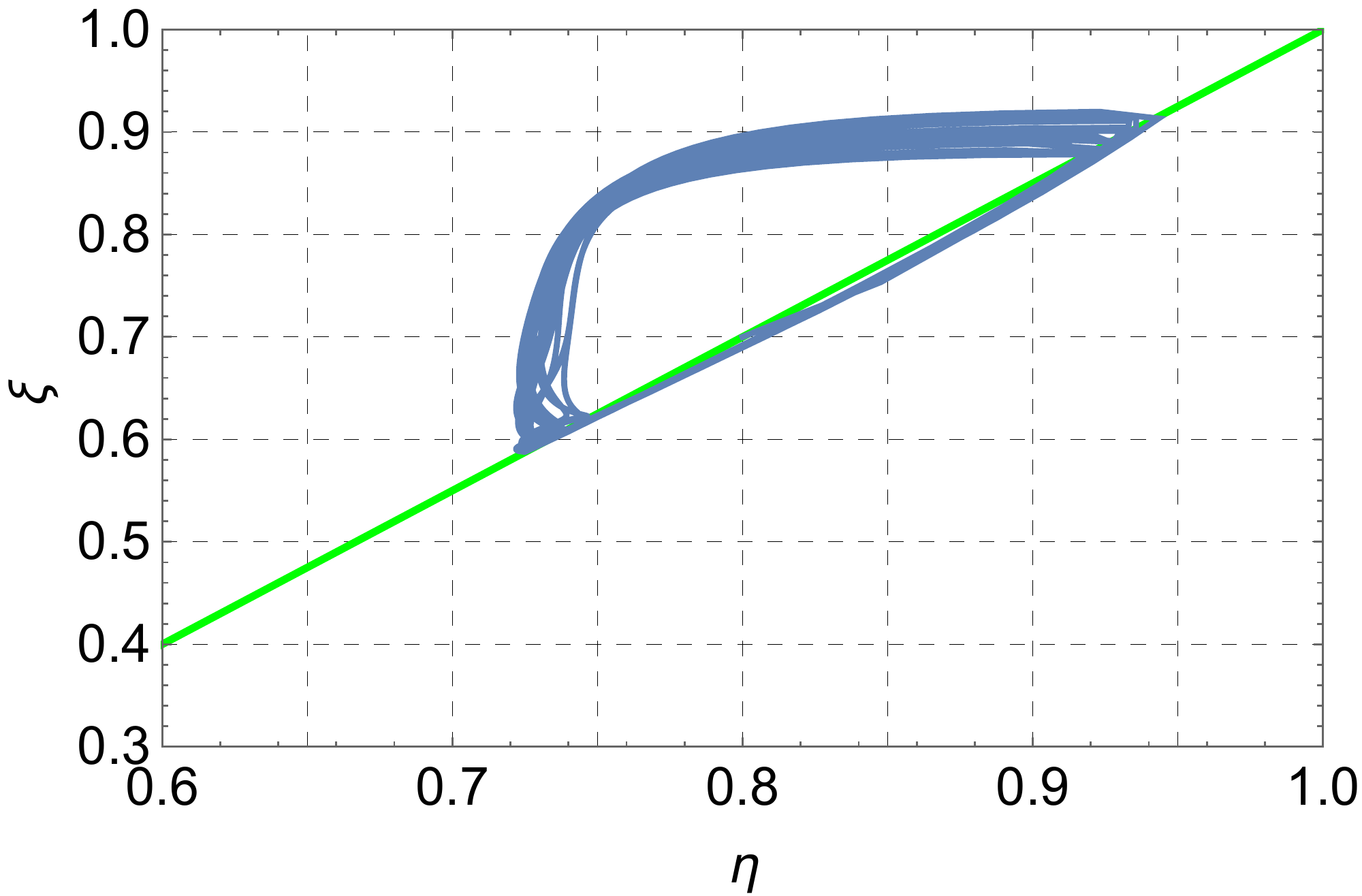}
\caption{Discontinuity plane cycles with full Milankovitch forcing}
\label{fig:13}
\end{subfigure}
\begin{subfigure}[b]{0.5\textwidth}
\includegraphics[width=\textwidth]{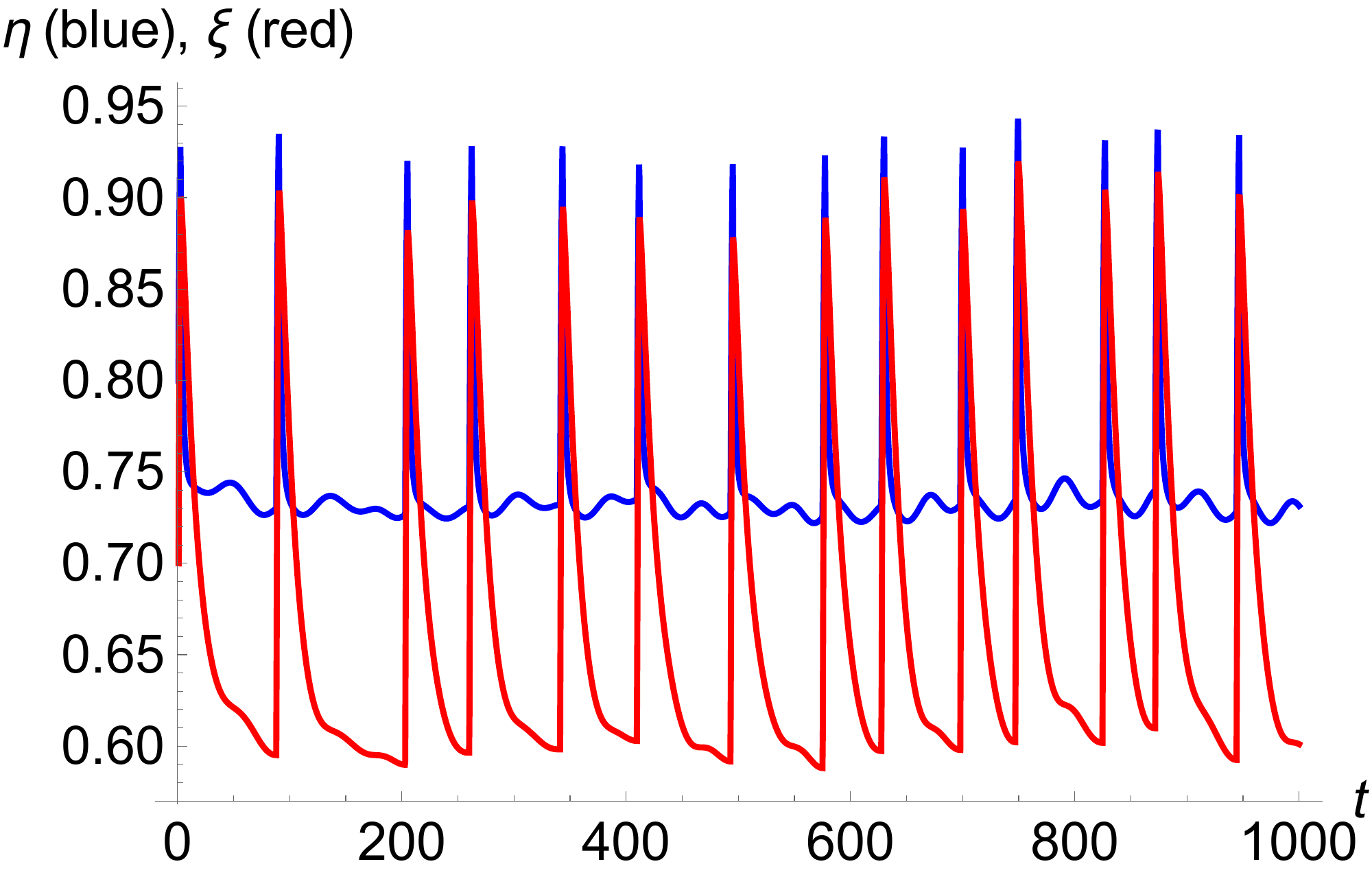}
\caption{Evolution of ice and snow lines over time with full Milankovitch forcing}
\label{fig:14}
\end{subfigure}
\caption{$b=b_0=2, b_1=45, a=1, \rho=\epsilon=4\times 10^{-2}$}
\end{figure}

\section{\fontsize{12}{13}\selectfont Discussion}
This unique model used was based on a finite approximation of an infinite dimensional model which comprised of Budyko's energy balance model, an ODE which describes the behavior of the edge of the ice sheet, and a snow line to account for glacial accumulation and ablation zones. Parameters such as insolation $Q$ and $s_2$ in the model were made to depend on Milankovitch cycles, that is, eccentricity of the Earth's orbit and the obliquity of the Earth's axis respectively and we observed that deglaciation and glaciation do occur mostly due to obliquity and to some extent eccentricity. Skipping of obliquity cycles was also seen which is in agreement with Huybers' model \cite{huybers2007glacial}. For further research, one may investigate how the inclusion of precession in this model affects the cycles and skipping. Thus, with this new simple conceptual model we were able to produce glacial cycles and force them with Milankovitch cycles that shed more light on which Milankovitch cycles factors were more dominant. 

\section*{\fontsize{12}{13}\selectfont Acknowledgements}
This project was supported by the University of Minnesota's Undergraduate Research Opportunities Program. I also would like to acknowledge the support of the Mathematics and Climate Research Network, which is partially supported by NSF Grants DMS-0940366 and
DMS-0940363. I appreciate the guidance given to me throughout this project by my mentor, Richard McGehee and also would like to recognize the support of his colleagues James Walsh, Esther Widiasih, and Jonathan Hahn. 

\bibliographystyle{plainurl}
\bibliography{MJUM}
\end{document}